\documentclass[reprint,amsmath,amssymb,aps,prl]{revtex4-1}

\usepackage{graphicx}
\usepackage{dcolumn}
\usepackage{bm}
\usepackage{hyperref}
\usepackage{epstopdf}
\makeatletter\renewcommand{\maketag@@@}[1]{\hbox{\m@th\normalsize\normalfont#1}}
\usepackage{color, soul, mathrsfs, url, titlesec}

\begin{document}

\preprint{APS/123-QED}

\title{A 15-user quantum secure direct communication network}
\author{Zhantong Qi$^{1,\dagger}$, Yuanhua Li$^{1,2,\dagger,*}$, Yiwen Huang$^{1,\dagger}$, Juan Feng$^{1}$, Yuanlin Zheng$^{1,3}$, Xianfeng Chen$^{1,3,4,*}$}

\affiliation{$^1$State Key Laboratory of Advanced Optical Communication Systems and Networks, School of Physics and Astronomy, Shanghai Jiao Tong University, Shanghai 200240, China
\\$^2$Department of Physics, Jiangxi Normal University, Nanchang 330022, China
\\$^3$Shanghai Research Center for Quantum Sciences, Shanghai 201315, China
\\$^4$Collaborative Innovation Center of Light Manipulation and Applications,
Shandong Normal University, Jinan 250358, China
\\$^{\dagger}$These authors contributed equally to this work
\\$^*$Corresponding authors
}

\begin{abstract}
	Quantum secure direct communication (QSDC) based on entanglement can directly transmit confidential information. However, the inability to simultaneously distinguish the four sets of encoded entangled states limits its practical application. Here, we explore a deterministic QSDC network based on time-energy entanglement and sum-frequency generation. 15 users are in a fully connected QSDC network, and the fidelity of the entangled state shared by any two users is greater than 97\%. The results show that when any two users are performing QSDC over 40 kilometers of optical fiber, the fidelity of the entangled state shared by them is still greater than 95\%, and the rate of information transmission can be maintained above 1Kbp/s. Our Letter demonstrates the feasibility of a proposed QSDC network, and hence lays the foundation for the realization of satellite-based long-distance and global QSDC in the future.
\end{abstract}
\maketitle

\phantomsection
\section{Introduction}
Quantum communication \cite{ref.1} has presented a revolutionary step due to its high security of the quantum information, which has been developed many models including quantum key distribution (QKD) \cite{ref.2}, quantum teleportation \cite{ref.3} and quantum secure direct communication (QSDC) \cite{ref.4}. Based on QKD technology, many different types of quantum communication networks have been proposed, such as an eight-user quantum communication network using multiple entangled states in the case of trusted node-free \cite{ref.5}, an entanglement-based wavelength-multiplexed quantum communication network \cite{ref.6}, and an integrated space-to-ground quantum communication network \cite{ref.7}. 
QSDC \cite{ref.8} sends secret information directly over a secure quantum channel. It does not require key distribution and key storage, and any method of attacking QSDC is to intercept the random number of the channel. Therefore, QSDC has very simple communication steps and security guarantees, which can extend the advantages of quantum communication between distant users. Recently, the experimental QSDC has developed significantly. QSDC protocols based on entanglement in Refs.\cite{ref.9, ref.10} have been experimentally realized by using quantum memory \cite{ref.11}, and fiber-photonics devices \cite{ref.12} respectively. Furthermore, QSDC based on entanglement can ensure the security against arbitrary eavesdropping attacks \cite{ref.13, ref.14}. However, the inability to simultaneously distinguish the four sets of encoded orthogonal entangled states limits its practical application, and a large-scale QSDC network cannot be realized. 
Here, we present a fully connected entanglement-based QSDC network including five subnets, with 15 users. Using the frequency correlations of the fifteen photon pairs via time division multiplexing (TDM) and dense wavelength division multiplexing (DWDM), we perform a 40-kilometer fiber QSDC experiment by implying two-step transmission between each user without generating secure keys. In this process, we divide the spectrum of the single-photon source into 30 International Telecommunication Union (ITU) channels. With these channels, there will be a coincidence event between each user by performing a Bell-state measurement (BSM) based on the sum-frequency generation (SFG). This allows the four sets of encoded entangled states to be identified simultaneously without post-selection, and the fidelity of the entangled photon pair after the SFG hardly changes, greater than 95\%. In our QSDC network, each user can request to communicate with others at any time. The connection relies on distributing entangled photon states between several users, expanding a wider range of applications for further quantum information processing processes.

\begin{figure}[htbp]
	\centering
	{\centerline{\includegraphics[width=7cm]{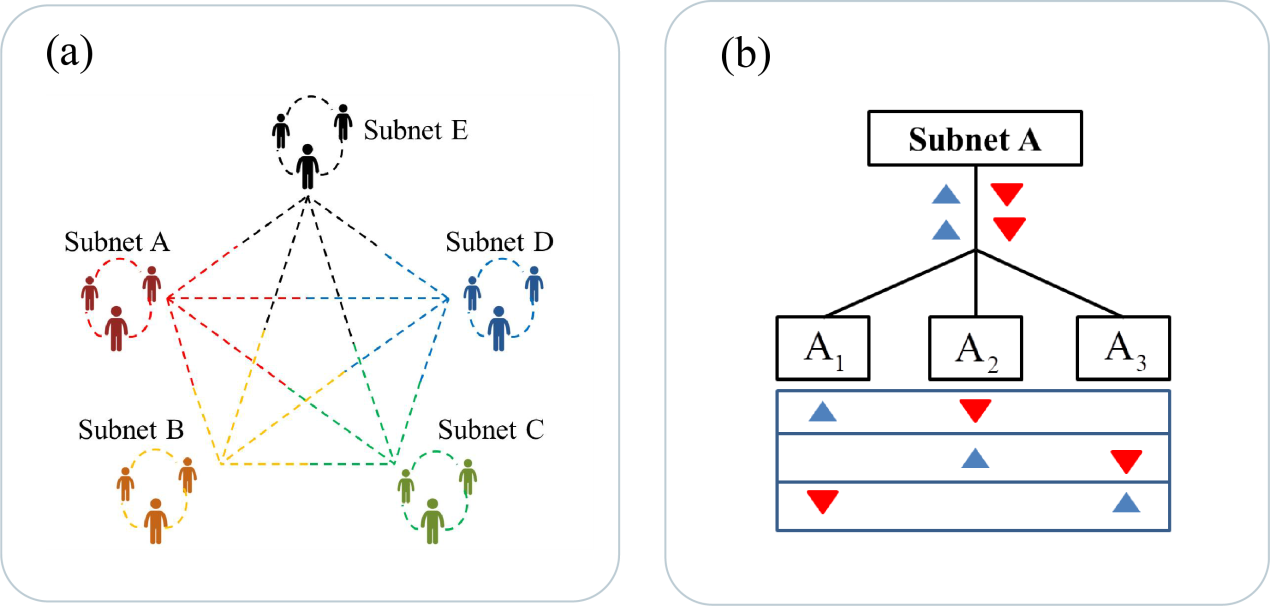}}}
	\caption{\label{Fig. 1}Composition of quantum network. (a) The quantum network is fully connected by five subnets (A, B, C, D and E are represented by red, orange, green, blue, and black, respectively). The dotted lines between the subnets (10 links with different colors) are the correlated time-energy photon pairs between the subnets. (b) Every subnet (such as subnet A) is equipped with a 1×3 BS and a delay controlling module, which splits a frequency-correlated entangled photon pair (red and blue signs) and sends them to three users randomly.}
\end{figure}

\phantomsection
\section{Network composition and experimental set-up}
We briefly describe the basic process of QSDC based on time-energy entanglement and SFG. We assume that any two users in the QSDC network want to communicate directly, i.e., User 1 wants to send information to User 2. They share N pairs of time-energy entangled states $\vert{\phi^+}\rangle=(\vert{ss}\rangle+\vert{ll}\rangle)/\sqrt{2}$ , where $s$ and $\l$ indicate that the entangled photons can either travel through a short or a long path. The detailed steps are as follows. (i) Detect the quantum channel to ensure its absolute safety. (ii) They agree that $\vert{\phi^+}\rangle, \vert{\phi^-}\rangle, \vert{\psi^+}\rangle$ and $\vert{\psi^-}\rangle$ encode the bit values 00, 01, 10, and 11, respectively. $\vert{\phi^\pm}\rangle=(\vert{ss}\rangle\pm\vert{ll}\rangle)/\sqrt{2}$  and $\vert{\psi^\pm}\rangle=(\vert{ls}\rangle\pm\vert{sl}\rangle)/\sqrt{2}$ are the four sets of Bell states. (iii) User 1 performs one of four unitary operations $I$ , $\sigma_z $, $\sigma_x $ , or $-i\sigma_y $  on the photons in his hand, to convert  $\vert{\phi^+}\rangle$ to $\vert{\phi^+}\rangle$ , $\vert{\phi^-}\rangle$ , $\vert{\psi^+}\rangle$, or $\vert{\psi^-}\rangle$ , respectively. These operations correspond to the encoding information 00, 01, 10, and 11, respectively. (iv) User 2 performs the BSM based on SFG to decode the information. This process allows the four sets of encoded Bell states to be identified simultaneously.

The complete network composition is divided into two layers as shown in Fig.~\ref{Fig. 1}. The left figure illustrates that the network processor provides 10 wavelength channels for combination between subnets, so that 10 links among the five subnets (A, B, C, D and E) constitute a fully connected quantum network. 10 time-energy entangled photon pairs between the subnets are divided into 20 ITU channels (1/-1 to 10/-10) via a 100 GHz DWDM. For connections within the subnet, quantum network processer will distribute the remaining 10 ITU wavelength channels (11/-11 to 15/-15). The resulting wavelength channels (that is, subnet A with ITU channels CH17-CH20) are sent to subnet A, which are used to communicate with subnets E, D, C and B, respectively (such as A and B via 4/-4, 11/-11 and 12/-12). Here, only six ITU channels corresponding to wavelength channel pairs are needed in the network to support the interconnection. The sharing of entangled photon pairs and interconnection between subnets are supported by controlling the distribution of four wavelengths from the entanglement source, so that each of them will have coincidence events with others. In each subnet, as shown in Fig.~\ref{Fig. 1}(b), a pair of entangled photons are separated by a passive beam splitter (BS) through TDM, and then randomly distributed to three users. For example, in the structure of subnet A, the upper triangle is the entangled photons of channel CH27, and the opposite triangle is the wavelength channel of CH37. In order to realize the connection among the three users, a time delay module will be controlled to achieve mutual coincidence events in the subnet. This way, the different wavelength allocation for 15 users can be realized by wavelength and TDM methods, in order to ensure the complete connection in the network.

\begin{figure*}[htbp]
	\centering
	{\centerline{\includegraphics[width=14.5cm]{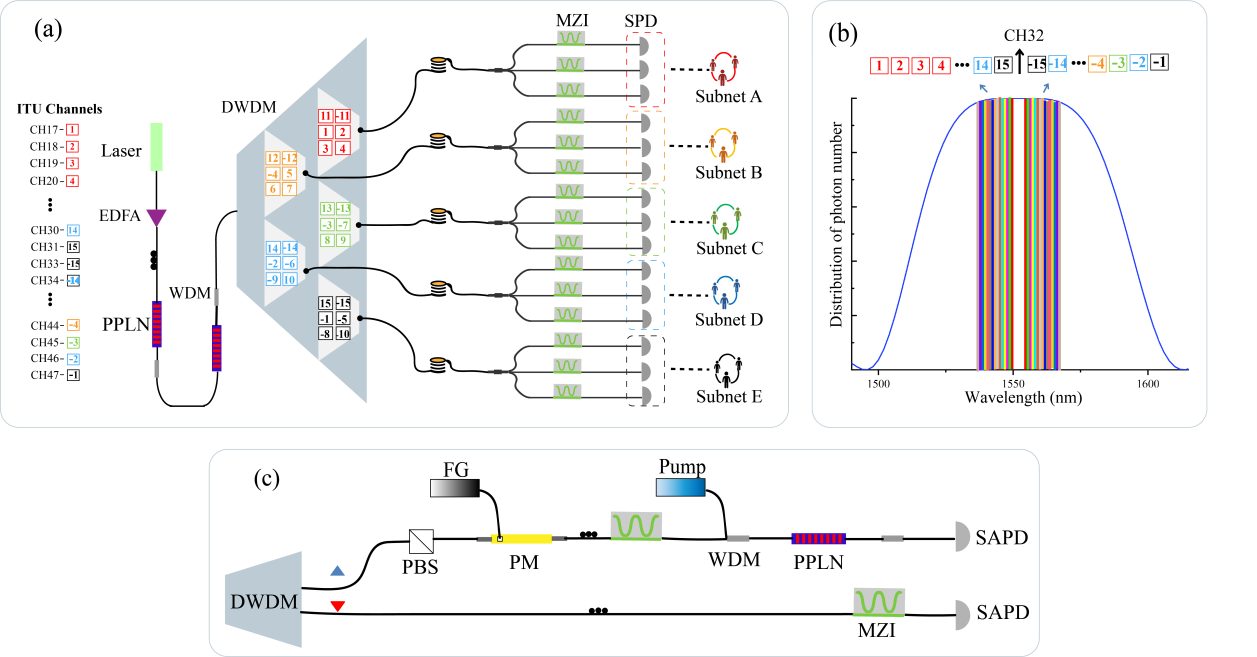}}}
	\caption{\label{Fig. 2} Experimental set-up. (a) The physical structure of the quantum network. The spectrum is split into 30 ITU grid channels via a 100 GHz-DWDM. CH17 to CH31 are numbered from 1 to 15 respectively, and the numbers with opposite sign denote corresponding to channels CH33-CH47. The architecture of wavelength allocation is omitted in the small trapezoid multiplex blocks. Each small WDM block with colored digital symbols constructs a wavelength group distributed by the network processor. (b) The spectrum of wavelength division multiplexing (WDM). The blue solid curve is the spectrum of the entangled photon source generated by SPDC process calculated by the Sellmeir equation \cite{ref.17}. Each pair of single photon and idler photon is indicated by the same colored bars with and without opposite digital sign. The y axis represents the number of photons corresponding to each channel. (c) Illustration of SFG progress. Photons generated in pairs by SPDC process are multiplexed into the SFG experiment to realize encoding and quantum communication.}
\end{figure*}

To generate SFG photons, we developed an entangled photon source with spontaneous parametric down-conversion (SPDC) process (supplementary material; Fig.~\ref{Fig. 2}(a)). In the experiment, we measured the fidelity of the generated entangled states by correlation records, larger than $97.5\%\pm1.0\%$. It is significant to confirm that the single photon source can provide high-quality photon pairs with all available ITU wavelength channels, as shown in Fig.~\ref{Fig. 2}(b), the spectrum of the single-photon source is divided into 30 ITU channels. 

Next, we will introduce the SFG encoding process between two users with silicon avalanche photodiodes (SAPDs) \cite{ref.15,ref.16}. In the experiment, we implement a frequency-correlated photon pair after SPDC process (CH34 and CH30) to realize the SFG process in the network (supplementary material; Fig.~\ref{Fig. 2}(c)). 

The QSDC process in network is performed in two steps with selected channels CH34 and CH30 (refer to Alice and Bob, respectively). We first establish a secure quantum entanglement channel over 40 kilometers of optical fiber. To guarantee the network security completely, single photons carrying polarization information are transmitted within the form of quantum data blocks. It is generally necessary to use quantum memory in the time domain to effectively control information transmission during QSDC tasks \cite{ref.11}, and the specific function is to adjust the time for storing photons. In our network, when the detection sequence is transmitted, we can achieve the function of quantum storage by adjusting the time delay of the electrical signal through a circuit delay module when the idler photons arrive at the detector. The delay can be adjusted according to the length of the detection sequence. After the safety detection is completed, the communication process is carried out. Firstly, user Alice selects a small part of photons distinguished in time as the detection sequence and sends them to Bob in quantum channel. Secondly, Bob has the option to randomly choose the $Z$-basis or $X$-basis to measure the single-photon received by Alice and then, publishes the position of measured photons, the measurement basis and measurement results through the classical synchronization channel. At the same time, additional wavelength channels can be alternatively blocked that are not needed for the communication. Thirdly, Alice chooses the same measurement basis as Bob to measure the corresponding entangled photons and then she compares the measurement results with the information informed by Bob and estimated the quantum bit error rates (QBERs), which can obtain the secrecy capacity of the system. Once the quantum bit error rate is lower than the threshold, after the comparison of the photon numbers between two users, the channel can be considered to be safe and unattended. In the second stage, the quantum network processer prepares $N$ Bell states with the form of $\vert{\phi^+}\rangle$ . The form of state contains the signal photons and idler photons from the entanglement source, and next the entangled photon sequences are separately sent to a pair of users (Alice with $S_A$ and Bob with $S_B$). Once Bob confirms the receipt of photon sequence ($S_B$), Alice applies different voltages to the polarization modulator \cite{ref.18} via a function generator (FG) and directly assign the information into the $S_A$  sequence with different voltages. Finally, Bob receives the sequence and then identifies the coded information from Alice through with or without SFG photons. In order to optimize the secure communication network, users can perform eavesdropping checking in the quantum channel at any time. If the monitored QBER is always lower than the threshold, it is considered that the communication is successful. In this case, each pair of legitimate users can perform block transmission in the second step, and repeat the process until the information transmission is completed.

\section{Results}

\begin{figure}[htbp]
	\centering
	{\centerline{\includegraphics[width=7.5cm]{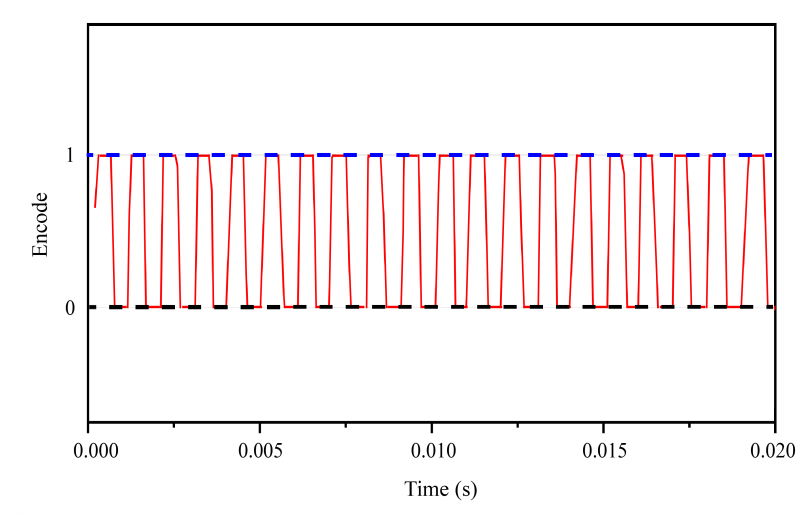}}}
	\caption{\label{Fig. 3}Channel security detection using single photons. The collected waveform (red) corresponding to the change with the generated SFG photon number is accumulated every 0.02 second while including a bin width of 300 ps. Codes ‘0’ and ‘1’ are indicated by the dotted blue and black lines respectively.}
\end{figure}

We performed the channel security detection in QSDC based on single photons\cite{ref.19}. The first step, as mentioned above, is to ensure the security for the transmission in the quantum channel, and then any two users in the network continue the connection step if the communication environment is demonstrated to be safe. Using the FG to apply a square wave voltage with the amplitude of 7 V, which operated at a frequency of 1 kHz, Alice modulates the polarization of the signal single photon (CH33) with a polarization modulator (PM). Bob receives the single photon after fiber-based transmission, and combines with a 1950 nm pump laser in order to perform up-conversion in the PPLN waveguide. The SAPD is digitized and deliver to the computer, where the photon counts value for coincidence events is used to reflect the result of encoding displayed on a waveform graph, as illustrated in Fig.~\ref{Fig. 3}. The maximum number of the generated SFG photons reaches $10^{5}$  per second. Consequently, the result obtained with the modulation of a single photon rotated from 0° to 90° will directly affect the generation number of SFG photons, which is reflected in the presence or absence of SFG photons. Assume that the assignment of ‘0’ or ‘1’, it is agreed with the higher and lower level of the waveform graph for clarity respectively. The collected waveform achieves a perfect performance in the absence of noise with the feedback of modulated information without mistakes. For observing the modulation result clearly, the speed of the PM modulation is adjusted to 1 kHz. However, during the practical detection process, users have an option to adjust the modulation rate via an FG, thereby matching the average number of generated SFG photons. In our experiment, the eavesdropper can only intercept part of the entangled particle in the system, and cannot obtain the overall state of the entangled quantum pair at the same time. That is to say, the action of Eve's eavesdropping only points to a random polarization state of the photon and limits the resolution of encoded operations, thus ensuring the security in the network. Furthermore, eavesdropping effect will be noticed by users through a significant decrease of the photon number.

\begin{figure}[htbp]
	\centering
	{\centerline{\includegraphics[width=8 cm]{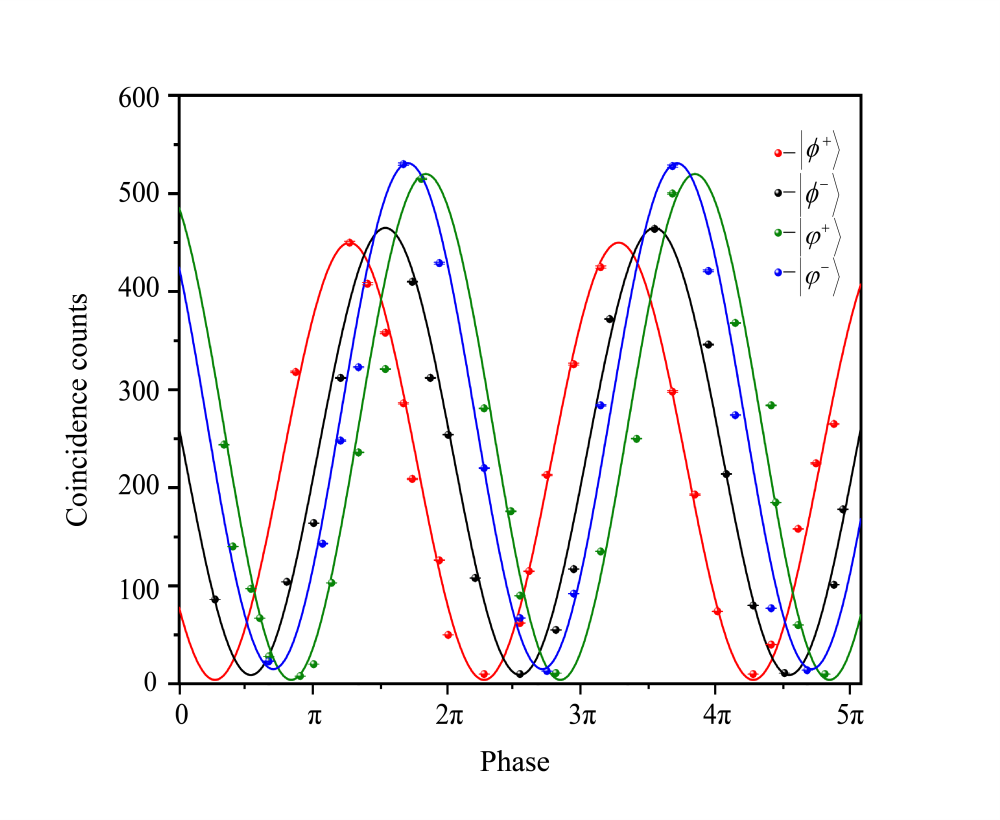}}}
	\caption{\label{Fig. 4} Experiment results. We measured two-photon interference fringe for four Bell states with subtraction of accidental coincidences. Note that the y axis represents the coincidence count varied with the full control of the phase over $2\pi$ rad. }
\end{figure}

After the detection, both users share the information when their detectors both clicked by detecting one photon within the coincidence window. In addition, we also conducted the fidelity measurement with four sets of Bell states, as shown in Fig.~\ref{Fig. 4}, and the measurement results show that the fidelity of four Bell states is almost as perfect as that before SFG, as listed in table.~\ref{table. 1}. Here, we can simultaneously distinguish the four Bell states by obtaining the maximum value of the number of generated SFG photons in the two-photon interference fringe. Still, the single photon characteristics after the SFG do not change.

\begin{table}[!htb]
\renewcommand\arraystretch{2}
\caption{\label{table. 1}Indication of the average fidelity with four Bell states.}
\setlength{\tabcolsep}{7mm}
\begin{tabular}{ccc}
\toprule 
Bell-state& Encode & Fidelity\\
\hline
$\vert{\phi^+}\rangle$ & 00 & $95.25\%\pm0.29\%$\\
$\vert{\phi^-}\rangle$ & 01 & $95.43\%\pm0.14\%$\\
$\vert{\psi^+}\rangle$ & 10 & $95.49\%\pm0.13\%$\\
$\vert{\psi^-}\rangle$ & 11 & $95.48\%\pm0.24\%$\\
\botrule
\end{tabular}
\end{table}

\phantomsection
\section{Discussion}
\textbf{Quantum network and users.}
The results obtained here show the characteristics of quantum network enabling the use of SFG process for communication. In what follows, we discuss some of aspects that are achieved for our experiment, as listed in the following three points. First, four Bell states can be resolved without post-selection when the SFG photons are detected, which is different from the linear optical BSM since it can only distinguish two Bell states. Second, by multiplexing the entangled states into one optical fiber and connecting to single photon detector for individual users, our design can directly provide convenience for several users to interconnect with any others. Users can optionally detect the photons that need for communication and block other wavelength signals without the involvement of the network provider. As the number of users increases, we can choose to increase ITU international wavelength channels or utilize narrower-band DWDMs. Third, according to the generation rate of SFG photons, we can achieve the faster transmission rate during network communication. It is important to comment that the improvement of the count rate for coincidence events is significantly related to high-performance of detector efficiency and a shorter coincidence window. Apart from sources of noise in the measurement, non-ideal aspects of FM such as modulation rate, the existence of unmodulated single photon and the response speed of the detector may contribute to limit the transmission rates and increase the accidental coincidence rates. Taking into account the total loss of the network architecture, we temporarily set 1 kHz rate for single-photon modulation in order to observe the modulate results. The main advantage is that the information transmission rate can be obtained based on the number of generated SFG photons. To generate better enhancement of the transmission rate, we can choose to accelerate the modulation speed to a level that matches the SFG photon generation rate (up to 100 Kbp/s). The alternative methods to increase the number of SFG photons are specific to improve the SFG efficiency and the quality of detectors.

\textbf{Security analysis.}
It is well known that the security and reliability of the information transmission for QSDC is an essential part in the quantum network \cite{ref.11}. Due to the imperfection efficiency of detectors and the loss of the transmission channel, insecure factors are likely to give the eavesdropper opportunities to intercept code information, namely the loss of the signal also means the leakage of information. Therefore, we implemented block transmission and step-by-step transmission methods in QSDC with estimating the secrecy capacity of the quantum channel. After confirming the security of the quantum channel, the legitimate user performs encoding or decoding operations within these schemes reliably, if not, they terminate the communication. In the following, we implemented the wiretap channel model \cite{ref.20, ref.21} to calculate the lower bound of the secrecy capacity $C_s$  (supplementary material)  \begin{equation}
	\begin{split}
		C_s \ge Q^B[1-H(e)]-Q^EH(e_x+e_z).
	\end{split}
	\label{eq:prediction}
\end{equation}     
Together with our encode method, our system yields a bit error rate of 0.0013, as shown in the extended data Fig. 2. According to the calculation using the above formula (1), the system can achieve almost perfect secrecy capacity transmission, which also illustrates that the security of information transmission is assured. More details for the security analysis of our quantum network can be found in the supplementary material.

\phantomsection
\section{Summary}
We have successfully realized a deterministic QSDC network based on entanglement and SFG with improved scalability. With this scheme, each user interconnects with any others through shared pairs of entangled photons in different wavelength. It is noteworthy that this approach can also be implemented for the multiple interconnection networks through setting quantum repeaters \cite{ref.22}. In addition, single-photon frequency conversion via cascaded nonlinear processes \cite{ref.23} can also be applied to achieve QSDC in large-scale distant networks. The concept of the quantum network design provides a framework for any realistic quantum communication systems, such as quantum teleportation. 

\phantomsection
\section{Acknowledgements}
This work is supported in part by the National Key Research and Development Program of China (Grant No. 2017YFA0303700), National Natural Science Foundation of China (Grant Nos. 11734011, 11804135, and 12074155), The Foundation for Shanghai Municipal Science and Technology Major Project (Grant No. 2019SHZDZX01-ZX06), and Project funded by China Postdoctoral Science Foundation (Grant No. 2019M661476).

\bibliographystyle{unsrt}
\bibliography{ref}

\end{document}